\input harvmac
\input epsf
\input amssym
%\draftmode
%
\noblackbox
%%%%%%%%%%%%%%%%%%%%%%%%%%%%%%%%%%%%%%%%%%%%%%
%%%%%%%%%%%%%%%%%%%%%%%%%%%
% some stuff needed for figures:
%%%%%%%%%%%%%%%%%%%%%%%%%%%%%%%%%%%%%%%%%%%%%%
%%%%%%%%%%%%%%%%%%%%%%%%%%%
\newcount\figno
\figno=0
\def\fig#1#2#3{
\par\begingroup\parindent=0pt\leftskip=1cm\rightskip=1cm\parindent=0pt
\baselineskip=11pt
\global\advance\figno by 1
\midinsert
\epsfxsize=#3
\centerline{\epsfbox{#2}}
\vskip -21pt
{\bf Fig.\ \the\figno: } #1\par
\endinsert\endgroup\par
}
\def\figlabel#1{\xdef#1{\the\figno}}
\def\encadremath#1{\vbox{\hrule\hbox{\vrule\kern8pt\vbox{\kern8pt
\hbox{$\displaystyle #1$}\kern8pt}
\kern8pt\vrule}\hrule}}
%%%%%%%%%%%%%%%%%%%%%%%%%%%%%%%%%%%%%%%%%%%%%%
%%%%%%%%%%%%%%%%%%%%%%%%%%%
% definitions
%%%%%%%%%%%%%%%%%%%%%%%%%%%%%%%%%%%%%%%%%%%%%%
%%%%%%%%%%%%%%%%%%%%%%%%%%%

\def\frac#1#2{{#1 \over #2}}

\def\semi{\subset\kern-1em\times\;}
\def\bar#1{\overline{#1}}
\def\sqr#1#2{{\vcenter{\vbox{\hrule height.#2pt
\hbox{\vrule width.#2pt height#1pt \kern#1pt \vrule width.#2pt}
\hrule height.#2pt}}}}

%

%

%

%

%

%
%%%%%%%%%%%%%%%%%%%%%%%%%%%%%%%%%%%%%%%%%%%%%%
%%%%%%%%%%%%%%%%%%%%%%%%%%%
% more definitions
%%%%%%%%%%%%%%%%%%%%%%%%%%%%%%%%%%%%%%%%%%%%%%
%%%%%%%%%%%%%%%%%%%%%%%%%%%

%
%\def\oneone{\rlap 1\mkern4mu{\rm l}}
%\def\coeff#1#2{\relax{\textstyle {#1 \over #2}}\displaystyle}

%%%%%%%%%%%%%%%%%%%%%%%%%%%%%%%%%%%%%%%%%%%%%%
%%%%%%%%%%%%%%%%%%%%%%%%%%%
% References
%%%%%%%%%%%%%%%%%%%%%%%%%%%%%%%%%%%%%%%%%%%%%%
%%%%%%%%%%%%%%%%%%%%%%%%%%%

\lref\curvcorr{A.~Dabholkar, ``Exact counting of black hole
microstates", [arXiv:hep-th/0409148]}

\lref\AdSBH{
J.~B.~Gutowski and H.~S.~Reall,
  ``General supersymmetric AdS(5) black holes,''
  JHEP {\bf 0404}, 048 (2004)
  [arXiv:hep-th/0401129].
  %%CITATION = HEP-TH 0401129;%%
}
%\BenaDE
\lref\BenaDE{
  I.~Bena and N.~P.~Warner,
  ``One ring to rule them all ... and in the darkness bind them?,''
  arXiv:hep-th/0408106.
  %%CITATION = HEP-TH 0408106;%%
}
%\ElvangDS
\lref\ElvangDS{H.~Elvang, R.~Emparan, D.~Mateos and H.~S.~Reall,``Supersymmetric black rings and three-charge supertubes,''
  Phys.\ Rev.\ D {\bf 71}, 024033 (2005)
  [arXiv:hep-th/0408120].
  %%CITATION = HEP-TH 0408120;%%
}
%\ElvangRT
\lref\ElvangRT{H.~Elvang, R.~Emparan, D.~Mateos and H.~S.~Reall,
``A supersymmetric black ring,''
  Phys.\ Rev.\ Lett.\  {\bf 93}, 211302 (2004)
  [arXiv:hep-th/0407065].
  %%CITATION = HEP-TH 0407065;%%
}

%\GauntlettQY
\lref\GauntlettQY{
  J.~P.~Gauntlett and J.~B.~Gutowski,
  ``General concentric black rings,''
  Phys.\ Rev.\ D {\bf 71}, 045002 (2005)
  [arXiv:hep-th/0408122].
  %%CITATION = HEP-TH 0408122;%%
}

%\ElvangSA
\lref\ElvangSA{
H.~Elvang, R.~Emparan, D.~Mateos and H.~S.~Reall,``Supersymmetric 4D rotating black holes from 5D black rings,''
  arXiv:hep-th/0504125.
  %%CITATION = HEP-TH 0504125;%%
}
%\BenaTK
\lref\BenaTK{
  I.~Bena and P.~Kraus,
  ``Microscopic description of black rings in AdS/CFT,''
  JHEP {\bf 0412}, 070 (2004)
  [arXiv:hep-th/0408186].
  %%CITATION = HEP-TH 0408186;%%
}
%\CyrierHJ
\lref\CyrierHJ{
  M.~Cyrier, M.~Guica, D.~Mateos and A.~Strominger,
  ``Microscopic entropy of the black ring,''
  arXiv:hep-th/0411187.
  %%CITATION = HEP-TH 0411187;%%
}

%\MaldacenaDE
\lref\MaldacenaDE{ J.~M.~Maldacena, A.~Strominger and E.~Witten,
``Black hole entropy in M-theory,'' JHEP {\bf 9712}, 002 (1997)
[arXiv:hep-th/9711053].
%%CITATION = HEP-TH 9711053;%%
}
%
%
%\BenaWV
\lref\BenaWV{
  I.~Bena,
  ``Splitting hairs of the three charge black hole,''
  Phys.\ Rev.\ D {\bf 70}, 105018 (2004)
  [arXiv:hep-th/0404073].
  %%CITATION = HEP-TH 0404073;%%
}
%\BenaNI
\lref\BenaNI{
  I.~Bena, P.~Kraus and N.~P.~Warner,
``Black Rings in Taub-NUT,''
  arXiv:hep-th/0504142.
  %%CITATION = HEP-TH 0504142;%%
}
%
%\GaiottoXT
\lref\GaiottoXT{
  D.~Gaiotto, A.~Strominger and X.~Yin,
  ``5D Black Rings and 4D Black Holes,''
  arXiv:hep-th/0504126.
  %%CITATION = HEP-TH 0504126;%%
}
\lref\Guica{M.~Guica, L.~Huang, W.~Li and A.~Strominger,
  ``$R^2$ Corrections for 5D Black Holes and Rings,''
  arXiv:hep-th/0505188.
  %%CITATION = HEP-TH 0505188;%%
}

%\BertoliniYA
\lref\Bert{ M.~Bertolini and M.~Trigiante, ``Microscopic
entropy of the most general four-dimensional BPS black  hole,''
JHEP {\bf 0010}, 002 (2000) [arXiv:hep-th/0008201].
%%CITATION = HEP-TH 0008201;%%
%\BertoliniEI
 ``Regular BPS
black  holes: Macroscopic and microscopic description of the
generating solution,''
Nucl.\ Phys.\ B {\bf 582}, 393 (2000) [arXiv:hep-th/0002191].
%%CITATION = HEP-TH 0002191;%%
}

%%%%%%%%%%%%%%%%%%%%%%%%%%%%%%%%%%%%%%%%%%%%%%
%%%%%%%%%%%%%%%%%%%%%%%%%%%
% Title
%%%%%%%%%%%%%%%%%%%%%%%%%%%%%%%%%%%%%%%%%%%%%%
%%%%%%%%%%%%%%%%%%%%%%%%%%%
\Title{
  \vbox{\baselineskip12pt \hbox{hep-th/0506015}
  \hbox{UCLA-05-TEP-16}
  \vskip-.5in}
}{\vbox{
  \centerline{$R^2 $ Corrections to Black Ring Entropy}
 }}

\centerline{Iosif Bena and Per Kraus}

\bigskip
\centerline{ \it Department of Physics and Astronomy,
UCLA, Los Angeles, CA 90095-1547, USA}

\medskip
\medskip
\medskip
\medskip
\medskip
\medskip
\baselineskip14pt
%%%%%%%%%%%%%%%%%%%%%%%%%%%%%%%%%%%%%%%%%
% Abstract
%%%%%%%%%%%%%%%%%%%%%%%%%%%%%%%%%%%%%%%%%
\noindent Recently, Guic\u a, Huang, Li, and Strominger considered
an $R^2$ correction to the entropy of a black ring, and found a
mismatch between supergravity and the CFT.  However, such a
comparison should take into account the subtle distinction between
the asymptotic charges of the black ring  and the charges entering
the CFT description. We show that using the correct charges yields
perfect agreement.

%%%
\Date{June, 2005}
%%%%%%%%%%%%%%%%%%%%%%%%%%%%%%%%%%%%%%%%%%%%%%
%%%%%%%%%%%%%%%%%%%%%%%%%%%
% Main text begins here
%%%%%%%%%%%%%%%%%%%%%%%%%%%%%%%%%%%%%%%%%%%%%%
%%%%%%%%%%%%%%%%%%%%%%%%%%%
\baselineskip14pt

\newsec{Introduction}

%\WaldNT
\lref\WaldNT{
  R.~M.~Wald,
  ``Black hole entropy is Noether charge,''
  Phys.\ Rev.\ D {\bf 48}, 3427 (1993)
  [arXiv:gr-qc/9307038].
  %%CITATION = GR-QC 9307038;%%
}

%\OoguriZV
\lref\OoguriZV{
  H.~Ooguri, A.~Strominger and C.~Vafa,
  ``Black hole attractors and the topological string,''
  Phys.\ Rev.\ D {\bf 70}, 106007 (2004)
  [arXiv:hep-th/0405146].
  %%CITATION = HEP-TH 0405146;%%
}

%\SenDP
\lref\SenDP{
  A.~Sen,
  ``How does a fundamental string stretch its horizon?,''
  arXiv:hep-th/0411255.
  %%CITATION = HEP-TH 0411255;%%
}

%\DabholkarBY
\lref\DabholkarBY{
  A.~Dabholkar, F.~Denef, G.~W.~Moore and B.~Pioline,
  ``Exact and asymptotic degeneracies of small black holes,''
  arXiv:hep-th/0502157.
  %%CITATION = HEP-TH 0502157;%%
}

%\HubenyJI
\lref\HubenyJI{
  V.~Hubeny, A.~Maloney and M.~Rangamani,
  ``String-corrected black holes,''
  arXiv:hep-th/0411272.
  %%CITATION = HEP-TH 0411272;%%
}

%\CardosoFP
\lref\CardosoFP{ G.~L.~Cardoso, B.~de Wit, J.~Kappeli and
T.~Mohaupt,
  ``Supersymmetric black hole solutions with $R^2$ interactions'',
[arXiv:hep-th/0003157];
  G.~Lopes Cardoso, B.~de Wit and T.~Mohaupt,
  ``Area law corrections from state counting and supergravity'',
  Class.\ Quant.\ Grav.\  {\bf 17}, 1007 (2000)
  [arXiv:hep-th/9910179];
  ``Macroscopic entropy formulae and non-holomorphic corrections for
  supersymmetric black holes'',
  Nucl.\ Phys.\ B {\bf 567}, 87 (2000)
  [arXiv:hep-th/9906094];
  ``Deviations from the area law for supersymmetric black holes'',
  Fortsch.\ Phys.\  {\bf 48}, 49 (2000)
  [arXiv:hep-th/9904005];
  ``Corrections to macroscopic supersymmetric black-hole entropy'',
  Phys.\ Lett.\ B {\bf 451}, 309 (1999)
  [arXiv:hep-th/9812082].
  %%CITATION = HEP-TH 9812082;%%
}

There has been a recent surge of interest in higher derivative
corrections to black hole entropy in string theory (an incomplete
list of references is
\refs{\CardosoFP,\OoguriZV,\curvcorr,\SenDP,\HubenyJI,\DabholkarBY}).
In certain cases it is possible to demonstrate precise agreement
between the supergravity and CFT corrections. A recent paper by
Guic\u a, Huang, Li, and Strominger \Guica\ considered this in the
context of five dimensional black rings
\refs{\ElvangRT,\BenaDE,\ElvangDS,\GauntlettQY}. On the
supergravity side they included an $R^2$ term known to be present
in M-theory on CY$_3$, and used Wald's formula \WaldNT\ to obtain
the entropy correction
\eqn\za{ \Delta S_{BR} = {\pi \over 6} c_2 \cdot p
\sqrt{{\hat{q}_0 \over D}}~.}
On the CFT side they took into account the correction to the central
charge, and obtained
\eqn\zb{\Delta S_{BR} = {\pi \over 6} c_2 \cdot p \sqrt{{\hat{q}_0
\over D}} +{\pi \over 24} c_2 \cdot p \sqrt{{D \over \hat{q}_0}}+
\ldots~.}

The second term represents a mismatch.  However, in performing
such comparisons one has to be careful to correctly translate the
charges of the black ring  into parameters appearing in the CFT.
The correct microscopic assignment appears in \BenaTK\  and differs
from the one that was discussed in \CyrierHJ\ and was used in \Guica\
to obtain equation \zb.
When the difference between the two  is taken into account, the
second term in \zb\ is absent, and the two sides agree, as we now
demonstrate.

\newsec{Two Candidates for the Microscopic Description of the Black Ring}

There exist two distinct proposals\foot{To facilitate easy
comparison, we follow the notation of \refs{\CyrierHJ,\Guica}. To
translate, use that $(Q_A,q_A)$ of \BenaTK\  are identified with
$(q_A,p^A)$ of \refs{\CyrierHJ,\Guica}.  There is also a flip in
the sign of $J_\phi$.} \refs{\BenaTK,\CyrierHJ} to count the
entropy of the 5D BPS black ring by relating it to the $(4,0)$ CFT
of the 4D BPS black hole \MaldacenaDE. The CFT entropy is
\eqn\sbh{S_{BR} = 2 \pi \sqrt{c_L \hat q_0 \over 6}}
and the leading contribution to the central charge is the product
of the dipole charges:
\eqn\central{c_L = 6 D = 6 D_{ABC} p^A p^B p^C~.}
However, the expression for $ \hat q_0$ is different. In \CyrierHJ\
is was argued that
\eqn\qw{\hat q_0 = -J_{\psi}+{1 \over 12 }D^{AB}q_A q_B + {c_L \over 24}}
where $J_{\psi}$ is the black ring angular momentum in the plane
of the ring, and the $q_A$ are the conserved charges of the black
ring as measured at infinity.

Earlier, in \BenaTK\ it had been argued (see pp. 5-6 of that
paper) that the black ring entropy is exactly that of a 4D black
hole with charges given by the seven parameters $ p^1, p^2,
p^3,\bar q_1 , \bar q_2 ,\bar q_3  $ and $J_{{\rm tube}} \equiv
J_{\psi}-J_{\phi} $. The microscopic entropy computation for
this 4D black hole was performed quite some time
ago \refs{\MaldacenaDE,\Bert}. The entropy
and central charge are given by \sbh\ and \central, except that
$\hat q_0$ is given by
\eqn\qr{\hat q_0' =-(J_{\psi}-J_{\phi}) + {1 \over 12 }D^{AB}\bar
q_A \bar q_B~. }
Written in this form, the black ring entropy is manifestly
$E_{7(7)}$ invariant  \BenaTK, as should be the case for a 4D
black hole entropy. By contrast, the  ${c_L \over 24}$ shift in
\qw\ obscures this invariance.
Refs. \refs{\BenaDE,\BenaTK}  argued that $J_{{\rm tube}}$ and
$\bar q_A$ are the momentum and charge living on the ring itself,
which differ from $J_\psi$ and $q_A$. The relation between the charges is
\eqn\qrel{\bar q_A  =  q_A - {6 D_{ABC} \over 2} p^B p^C~. }
As one can easily check, $\hat q_0$ and $\hat q_0'$ are equal, so
they both yield agreement at leading order with the entropy
calculated in gravity \refs{\BenaDE,\ElvangDS}.

This situation resulted in some confusion concerning which charges
should appear in the CFT. Note that both \BenaTK\ and \CyrierHJ\
were trying to replace the circular black ring with a straight
black string, which is a discontinuous process and so inherently
ambiguous.

Fortunately, it is possible to resolve this puzzle unambiguously
by  putting the black ring in Taub-NUT, implementing the
philosophy suggested and used in \GaiottoXT. In \BenaNI\ it was
shown that by adjusting the moduli it is possible to move the
black ring arbitrarily far from the Taub-NUT center,  and the
black ring solution then reduces to the 4D black hole/5D black
string solution constructed in \BenaWV. One finds that the
momentum and  charges of this 4D black hole are $J_{{\rm tube}}$
and $\bar q_A $. In this way one {\it derives} the 4D black hole
description rather than having to assume it, and one has no choice
in the charge assignments.
 This interpolation thus proves that the correct microscopic
description of the black ring is \qr\ and not \qw.

\newsec{$R^2$ corrections}

We now proceed to show that by using the microscopic
description \qr\ we resolve the mismatch \Guica\ between the $R^2$
corrections to the gravitational entropy and the corresponding
corrections in the CFT.

The corrections on the $CFT$ side are obtained by considering the
first correction to $c_L$:
\eqn\central{c_L  \rightarrow c_L + \Delta c_L}
where $ \Delta c_L = c_2 \cdot p$. If one uses the microscopic
description \qw, a correction to $c_L$ affects both the explicit
$c_L$ in \sbh\ as well as  $\hat q_0$. The change in the
microscopic entropy is given by equation (11) in \Guica:
\eqn\eleven{\Delta S_{BR} = {\pi \over 6} c_2 \cdot p \sqrt{{\hat
q_0 \over D }}+ {\pi \over 24} c_2 \cdot p \sqrt{{D \over \hat q_0
}}+... }

However, if one uses the microscopic description \qr, the
correction to $c_L$ does not affect $\hat{q}_0'$, and the change
in the entropy formula is
\eqn\elevena{\Delta' S_{BR} = {\pi \over 6} c_2 \cdot p
\sqrt{{\hat q_0' \over D }}+ ... }
This agrees exactly with the correction coming from the $R^2$
terms in the gravity description \za\ (equation (9) in \Guica).
Thus in the microscopic description \qr\ the second term in
\eleven\  simply does not exist, and the agreement with gravity is
perfect. This gives further confirmation for  the correctness of
the microscopic description \qr.\foot{One might try to argue that
the second term in equation \eleven\ is subleading and therefore
not captured by the Cardy formula. However, this argument could
equally well be applied to the terms of the order $(p^1 p^2
p^3)^2$ in the uncorrected entropy. But omission of such terms
would destroy the match between the supergravity and CFT entropies
at the leading order.  }

\bigskip
\noindent {\bf Acknowledgements:} \medskip \noindent The work of
IB and PK is supported in part by NSF grant 0099590.

\listrefs
\end